# Resonantly Trapped Bound State in the Continuum Laser

T. Lepetit*, Q. Gu*, A. Kodigala*, B. Bahari, Y. Fainman, B. Kanté

*Department of Electrical and Computer Engineering, University of California San Diego, La Jolla, CA 92093-0407, USA*

bkante@ucsd.edu

*These authors contributed equally to this work

**Abstract**- Cavities play a fundamental role in wave phenomena from quantum mechanics to electromagnetism and dictate the spatiotemporal physics of lasers. In general, they are constructed by closing all "doors" through which waves can escape. We report, at room temperature, a bound state in the continuum laser that harnesses optical modes residing in the radiation continuum but nonetheless may possess arbitrarily high quality factors. These counterintuitive cavities are based on resonantly trapped symmetry-compatible modes that destructively interfere. Our experimental demonstration opens exciting avenues towards coherent sources with intriguing topological properties for optical trapping, biological imaging, and quantum communication.

In 1929, only three years after the advent of quantum mechanics, von Neumann and Wigner showed for the first time that Schrödinger's equation can have bound states above the continuum threshold [1]. These peculiar states, called bound states in the continuum (BICs), manifest themselves as resonances that do not decay. For several decades after, the idea lay dormant in large part because it was regarded primarily as a mathematical curiosity. In 1977, Herrick and Stillinger revived interest in BICs when they suggested that BICs could be observed in semiconductor superlattices [2,3]. More than a decade later, in 1992, BICs were observed experimentally in $Al_xIn_{1-x}As/Ga_yIn_{1-y}As$ superlattices [4]. BICs naturally arise from Feshbach's quantum mechanical theory of resonances, as explained by Friedrich and Wintgen, and are thus more physical than initially realized [5]. Recently, it was understood that BICs are intrinsically a wave phenomenon and are not restricted to the realm of quantum mechanics. They have since been shown to occur in many different fields of wave physics including acoustics, microwaves, and nanophotonics [6-14]. However, experimental observations of bound states in the continuum have been limited to passive systems.

Here, we report, at room temperature, the first lasing action from a bound state in the continuum cavity. Our results show that the lasing wavelength of fabricated BIC cavities, made of arrays of cylindrical nanoresonators suspended in air, scales with nanoresonators radii according to the theoretical prediction for the BIC mode. Moreover, lasing action from the designed BIC cavity persists even after scaling down the membrane to as few as 8-by-8 nanoresonators. BIC lasers open new avenues in the study of light-matter interaction as they are intrinsically connected to topological charges [15], and, represent natural vector beam sources [16], which are highly sought after in the fields of optical trapping, biological sensing, and quantum information.

Generally, open systems are described by non-Hermitian effective Hamiltonians that have multivariate and complex eigenvalues describing modes of the system. These eigenvalues exist in a multidimensional space (hyperspace), but in a given frequency range, the investigation can be reduced to a finite number of variables, limiting the complexity of the effective Hamiltonian [7]. When eigenvalues come close to crossing as a function of a geometrical parameter that modifies the system, avoided resonance crossing (ARC) occurs, i.e., eigenvalues repel each other in the entire complex plane [17-18]. Friedrich and Wintgen showed that resonantly trapped BICs represent a particular type of ARC for which coupling occurs predominantly in the far field.

Our system consists of a thin membrane of semiconductor material suspended in air. The field in the air is a superposition of plane waves, which are interpreted as independent decay channels, and can be either propagating or evanescent. We subsequently structure the membrane at the nanoscale. The field in the membrane, which becomes a superposition of coupled plane waves due to structuring, is also coupled to the field in air. In the resulting open system described by a non-Hermitian Hamiltonian, resonance lifetime is governed by coupling amongst different channels. The imaginary part of the complex frequency serves to quantify the decay of modes. BICs arise when complex frequency modes (in the continuum) interfere destructively to give a purely real frequency mode. They are very peculiar discrete modes in that they are actually embedded within the continuous spectrum but intrinsically possess an infinitely high radiation quality factor as a result of their non-decaying nature. BICs are thus ideally suited for the design of perfect nanophotonic cavities.

As shown in Fig. 1, our BIC cavity is composed of a periodic array of nanoresonators of radius $R$ interconnected by a network of supporting bridges used for the mechanical stability of the system. The membrane consists of several $In_xGa_{1-x}As_yP_{1-y}$ multiple quantum wells, specially designed to operate around telecommunication wavelengths ($\lambda \approx 1.55$ μm). The radius of the nanocylinders is the only parameter we use to tune the modes of the membrane and alter the effective Hamiltonian. The structure is fabricated using electron-beam lithography and reactive ion etching (RIE) to define the cylindrical resonators followed by a wet etching step to create the membrane. It is worth noting that the radius of the fabricated nanoresonators is always smaller than their nominal design values, a consequence of RIE. As a result, the maximum achievable radius will be smaller than $p/2$ where $p$ is the period of our structure.

To analyze our system, we calculate the quality factors at normal incidence around 1.55 μm where the material gain peaks. We restrict the discussion to odd modes (transverse magnetic like) as they have much higher quality factors than even modes (transverse electric like) in the frequency range of interest. We find three modes around 1.55 μm with appreciable quality factors, one doubly degenerate mode (modes 1-2) and one singly degenerate mode (mode 3). Figure 2A shows their quality factor as a function of the radius (530 nm ≤ $R$ ≤ 542 nm). The quality factor of mode 3 is independent of the radius and remains high throughout the calculated range. This mode corresponds to a symmetry-protected mode [19]. In contrast, the quality factor of modes 1-2 strongly depends on the radius and reaches a maximum at an optimum radius of $R_{opt}$ = 536 nm. At this optimum radius, modes 1-2 completely decouple from the radiation continuum and thus become BICs. Indeed, the quality factor can diverge in two situations depending on whether we are considering an isolated resonance or trapped resonances. In the first situation (isolated resonances, mode 3), coupling to the outside vanishes solely as a result of symmetry protection. Any perturbation that preserves symmetry, such as a modification of the radius, has no impact on its quality factor. This type of mode has been extensively studied before [20]. In the second situation (trapped resonances modes 1-2), coupling to the outside vanishes as a result of destructive interference [13,21]. Resonantly trapped BICs achieve an infinite quality factor at the singular radius $R_{opt}$ but the quality factor remains very high for radii around $R_{opt}$. Figure 2B shows the transmission spectrum at normal incidence of our structure in which the infinite quality factor of modes 1-2 can be seen from the vanishing linewidth.

Figure 2C shows the dispersion relation of the BIC structure at $R = R_{opt}$ along MΓ and ΓX. We plot the complex dispersion relation of modes 1-2 (Fig. 2D-E) and mode 3 (Fig. 2F). Figure 2F shows that mode 3 is extremely sensitive to symmetry-breaking perturbations as its quality factor drops sharply away from the Γ point. Quality factors of modes 1-2, which are no longer degenerate away from the Γ point (as seen in Fig. 2C), do not drop as sharply as that of mode 3. Modes 1-2 are thus much less sensitive to symmetry-breaking perturbations. Additionally, the resonance-trapped BIC is robust because a variation in radius only induces its displacement in k-space whereas a symmetry-breaking perturbation destroys the symmetry-protected mode [13]. This is of utmost importance in device design as fabrication tolerances will have less impact on resonance-trapped BIC than on modes that rely on symmetry protection. Moreover, designing a mode with a high quality factor in a large region of k-space is of practical importance because fabricated devices, which are never infinite, always sample the dispersion relation in a finite neighborhood in k-space [22]. Therefore, for a given quality factor, we can achieve a much smaller device footprint with a resonance-trapped BIC mode than with symmetry-protected modes.

To experimentally demonstrate lasing from our BIC cavity, we optically pump the membrane at room temperature with a pulsed laser (λ = 1064 nm, T = 12 ns pulse at f = 300 kHz repetition rate) and record the spectral emission. Figure 3A shows the evolution of the output power as a function of both the pump power and the wavelength. At low pump power, we observe a spectrally broad photoluminescence spectrum, while at high pump power, we observe a drastic overall suppression of the photoluminescence in favor of one extremely narrow peak, *i.e.*, lasing. As depicted in Figure 3A, three modes show amplification at first ($P_{pump} ≈ 60$ μW) but, ultimately, only one remains ($P_{pump} ≈ 120$ μW). Lasing action occurs at a wavelength of 1553.2 nm with a linewidth of 0.52 nm (see inset of Figure 3B). Figure 3B shows the evolution of the output power as a function of the pump power around this lasing wavelength. We observe a clear threshold behavior with a threshold power of 62 μW or a density of 108 mW.mm$^{-2}$.

To further demonstrate the robustness and scalability [23] of our BIC laser, we fabricated several devices with a range of radii and array sizes. Figure 4A, shows the measured lasing wavelength of devices of different array size (8x8, 10x10, 16x16, and 20x20), and different radii of nanoresonators. The solid and dashed lines represent respectively the theoretical resonant wavelength of modes 1-2 and 3 for different radii of nanoresonators for the infinite array. We observe a very good agreement between the experimental lasing wavelengths and the theoretical resonant wavelengths of the resonance-trapped BIC mode (mode 1-2). This agreement confirms that lasing action is indeed from the BIC mode over the entire range of radii. Moreover, the persistence of lasing for all array sizes down to as few as 8-by-8 resonators shows the scalability of our BIC laser, thanks to the large quality factor of the resonance-trapped BIC mode in a wide region of k-space.

We reported the first bound state in the continuum (BIC) laser from a cavity that can surprisingly have arbitrarily high quality factors despite being embedded in the continuum of radiation modes. Our cavity, made of an array of suspended cylindrical nanoresonators, shows persistent single mode lasing for various radii and array sizes. The lasing wavelength follows the theoretical prediction of the BIC mode. These results demonstrate the robustness and scalability of the system. The ability to confine light within the radiation continuum opens the door to the study of the intriguing topological physics of BICs and the realization of non-standard photonic devices, sensors, and sources.

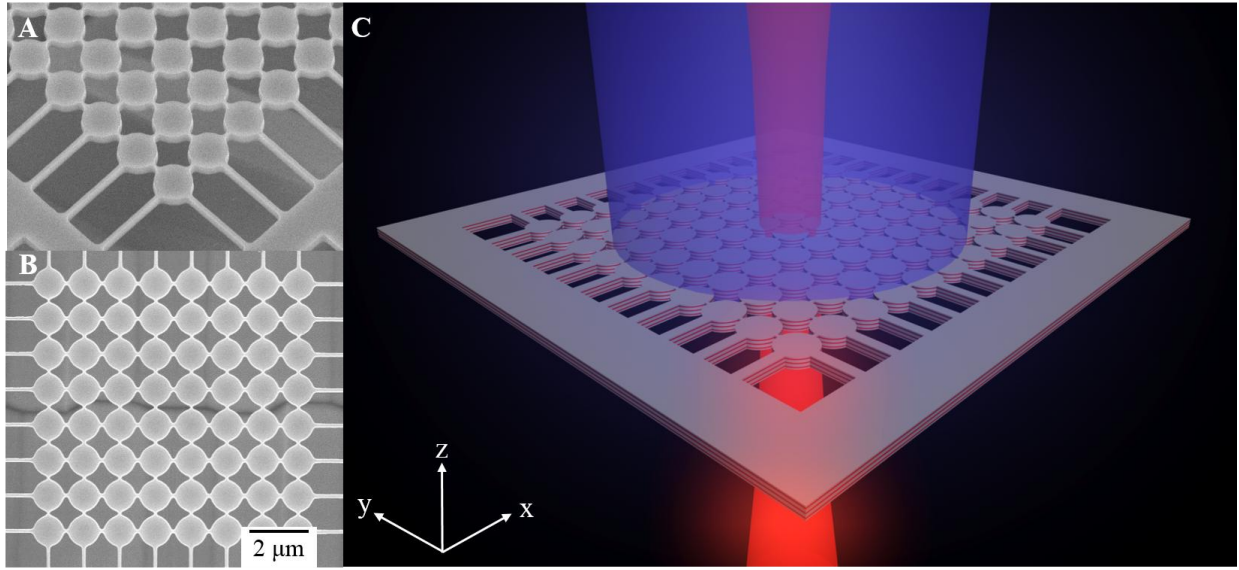

**Fig. 1. Bound state in the continuum laser.** (**A**) Tilted electron micrograph of InGaAsP multiple quantum wells cylindrical nanoresonator array suspended in air. All structures are fabricated using electron beam lithography followed by reactive ion etching to form the cylinders. We subsequently use wet etching to suspend the structure. (**B**) Top view of an 8-by-8 array with supporting bridges, which are used for the mechanical stability of the membrane. The dimensions of the structure are: *period* = 1200 nm, *thickness* = 300 nm, and *bridge width* = 100 nm. (**C**) Schematic of the fabricated system illustrating the larger pump beam (blue) and lasing from the bound state in the continuum (BIC) mode (red). The radius of the nanocylinders is the key parameter in our BIC design.

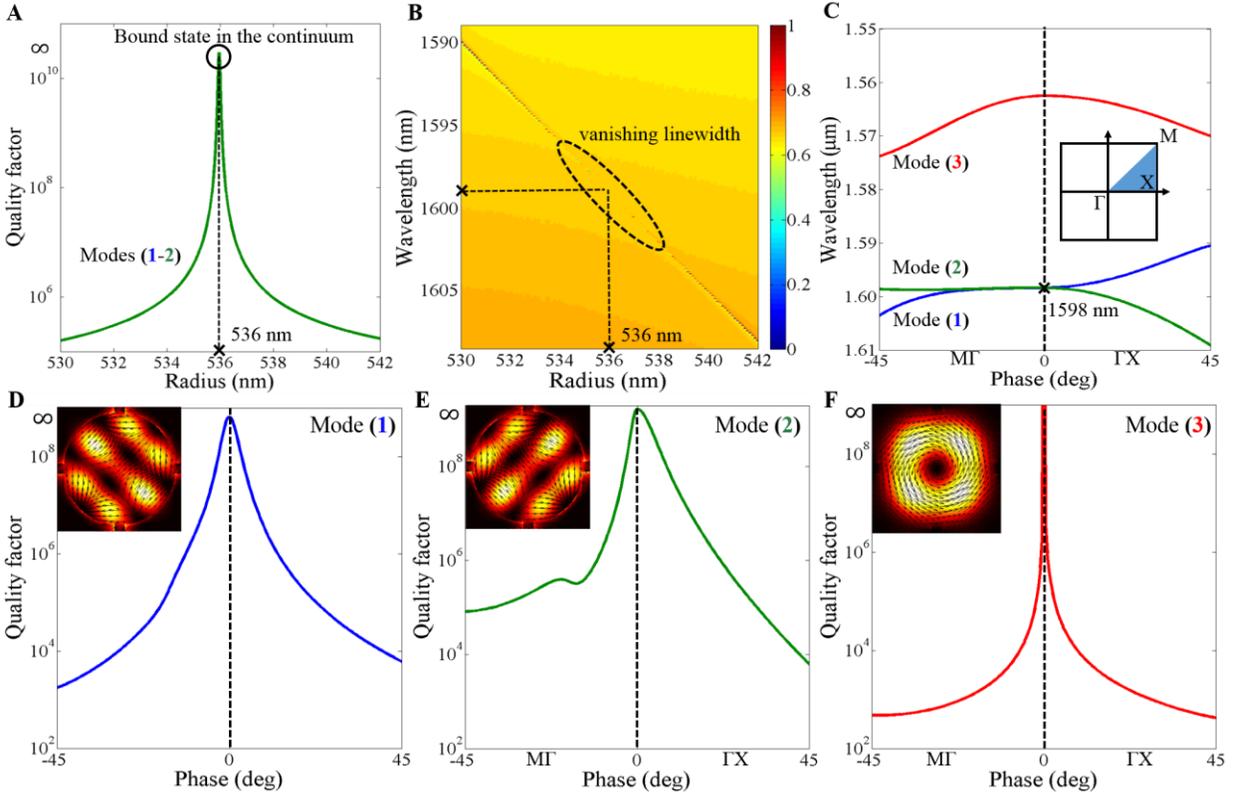

**Fig. 2. Design and complex dispersion relation of the bound state in the continuum cavity.** (**A**) Quality factor of high-Q modes at the Γ point around telecommunication wavelength for different nanoresonator radii. The quality factors of modes 1-2, which are doubly degenerate at the Γ point, are strongly dependent on the radius $R$. For $R = R_{opt}$ at 536 nm, this quality factor approaches infinity to form a resonantly trapped bound state in the continuum. For radii around $R_{opt}$ the quality factor remains very high. (**B**) Transmission spectrum at normal incidence showing the vanishing linewidth of modes 1-2 when the radius approaches $R_{opt}$. The wavelength of the modes is a function of the radius and it continuously varies between the smallest radius ($R$ = 530 nm) and largest radius ($R$ = 542 nm). (**C**) Dispersion relation around 1.55 μm for high-Q modes (1, 2 and 3) in both MΓ and ΓX directions. The inset shows the first Brillouin zone of the square lattice and irreducible contour for cylindrical nanoresonators (shaded area). The contour connects high-symmetry points Γ, X, and M. Quality factor of high-Q modes in both MΓ and ΓX directions for mode 1 (**D**), mode 2 (**E**), and mode 3 (**F**). Insets represent the normalized electric field on the surface of the cylinder. Modes 1 and 2 are identical under 90-degree rotation. Mode 3 is a symmetry protected mode and is thus not affected by geometrical changes that preserve symmetry, such as the change of radius. The quality factor of mode 3, however, drops rapidly away from the high symmetry point Γ. It drops more rapidly compared to the quality factor of mode 1-2. The sharper drop of the quality factor of mode 3 away from Γ implies that, the integrated quality factor of this mode will be smaller than those of modes 1 and 2 in the case of finite sized samples.

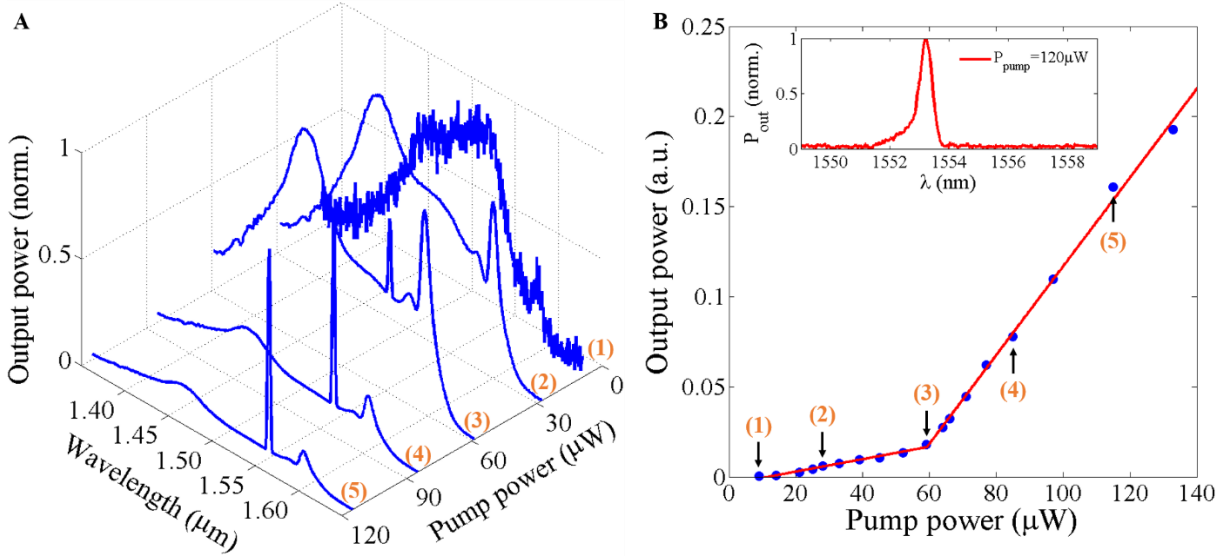

**Fig. 3. Experimental characterization of bound state in the continuum laser.** (**A**) Evolution of the normalized output power as a function of both wavelength (μm) and pump power (μW) for a 20-by-20 array with a nanoresonator radius of 507 nm. We observe the transition from a broad spontaneous emission to a single lasing peak at 1553.2 nm. (**B**) Output power as a function of the pump power (light-light curve) around the lasing wavelength. We observe the onset of lasing at a threshold power of 62 μW. The blue dots correspond to measurements and numbers 1 through 5 denote spectra plotted in (A). The inset shows the lasing spectrum at a pump power of 120 μW (number 5) with a linewidth of 0.52 nm.

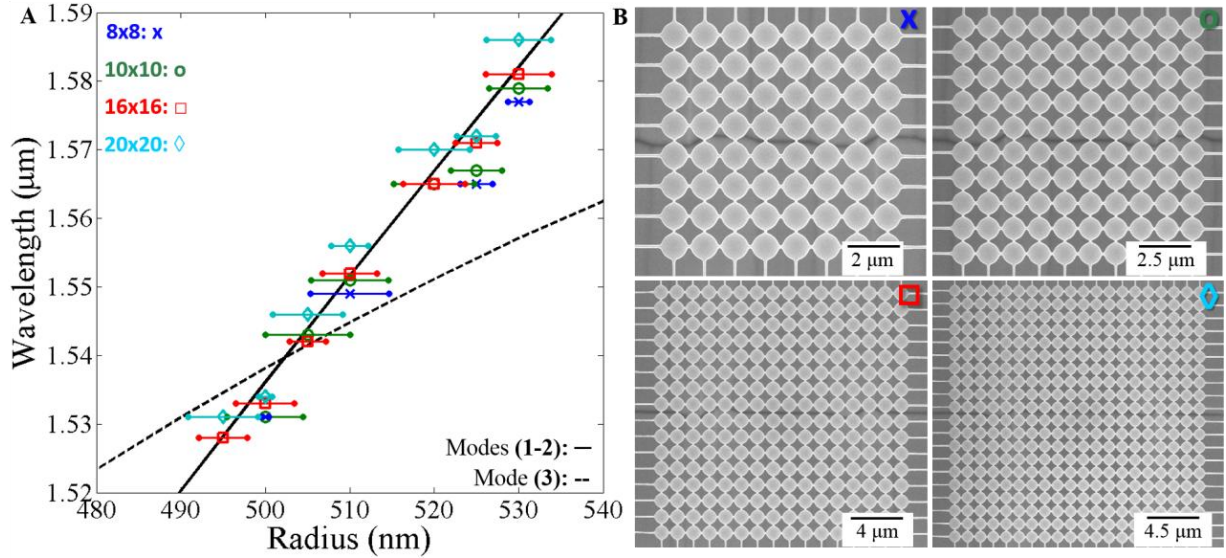

**Fig. 4. Scaling of bound state in the continuum lasers.** (**A**) Lasing wavelength as a function of nanoresonator radius, from 495 nm to 530 nm, with array sizes of 8-by-8 (cross), 10-by-10 (circle), 16-by-16 (square), and 20-by-20 (diamond). Each point corresponds to a device with a specific radius and array size. Error bars indicate the standard deviation of radii measured from fabricated devices. The lines represent the theoretical resonant wavelength of modes 1-2 (solid line) and 3 (dashed line) for different radii of nanoresonators, for the infinite array. The good agreement between the experimental lasing wavelengths and the theoretical resonant wavelengths of the resonance-trapped BIC mode (mode 1-2) confirms that lasing action is indeed from the BIC mode. (**B**) Electron micrographs of fabricated BIC lasers of size 8-by-8, 10-by-10, 16-by-16, and 20-by-20.

**References:**

[1] J. von Neumann and E. Wigner, "On some peculiar discrete eigenvalues", *Phys. Z* **30**, 467 (1929).

[2] D. R. Herrick, "Construction of bound states in the continuum for epitaxial heterostructure superlattices", *Physica B* **85**, 44-50 (1977).

[3] F. H. Stillinger, "Potentials supporting positive-energy eigenstates and their application to semiconductor heterostructures", *Physica B* **85**, 270-276 (1977).

[4] F. Capasso *et al.*, "Observation of an electronic bound state above a potential well", *Nature* **358**, 565-567 (1992).

[5] H. Friedrich and D. Wintgen, "Interfering resonances and bound states in the continuum", *Phys. Rev. A* **32**, 3231 (1985).

[6] C. Linton *et al.*, "Embedded trapped modes in water waves and acoustics", *Wave Motion* **45**, 16 (2007).

[7] T. Lepetit and B. Kanté, "Controlling multipolar radiation with symmetries for electromagnetic bound states in the continuum", *Phys. Rev. B* **90**, 241103(R) (2014).

[8] D. C. Marinica *et al.*, "Bound states in the continuum in photonics", *Phys. Rev. Lett.* **100**, 183902 (2008).

[9] E. N. Bulgakov and A. F. Sadreev, "Bound states in the continuum in photonic waveguides inspired by defects", *Phys. Rev. B* **78**, 075105 (2008).

[10] F. Dreisow *et al.*, "Adiabatic transfer of light via a continuum in optical waveguides", *Opt. Lett.* **34**, 2405-2407 (2009).

[11] Y. Plotnik *et al.*, "Experimental observation of optical bound states in the continuum", *Phys. Rev. Lett.* **107**, 183901 (2011).

[12] S. Weimann *et al.*, "Compact surface Fano states embedded in the continuum of waveguide arrays", *Phys. Rev. Lett.* **111**, 240403 (2013).

[13] C. W. Hsu *et al.*, "Observation of trapped light within the radiation continuum", *Nature* **499**, 188-191 (2013).

[14] F. Monticone and A. Alù, "Embedded photonic eigenvalues in 3D nanostructures", *Phys. Rev. Lett.* **112**, 213903 (2014).

[15] B. Zhen *et al.*, "Topological nature of optical bound states in the continuum", *Phys. Rev. Lett.* **113**, 257401 (2014).

[16] E. Miyai *et al.*, "Photonics: Lasers producing tailored beams", *Nature* **441**, 946 (2006).

[17] J. Wiersig, "Formation of long-lived, scarlike modes near avoided resonance crossings in optical microcavities", Phys. Rev. Lett. **97**, 253901 (2006).

[18] E. Persson *et al.*, "Observation of resonance trapping in an open microwave cavity", Phys. Rev. Lett. **85**, 2478 (2000).

[19] K. Sakoda, Optical Properties of Photonic Crystals, 2005.

[20] S. Fan and J. D. Joannopoulos, "Analysis of guided resonances in photonic crystal slabs", *Phys. Rev. B* **65**, 235112 (2002).

[21] Y. Yang *et al.*, "Analytical perspective for bound states in the continuum in photonic crystal slabs", *Phys. Rev. Lett.* **113**, 037401 (2014).

[22] T. Xu *et al.*, "Confined modes in finite-size photonic crystals", *Phys. Rev. B* **72**, 045126 (2005).

[23] A. F. Koenderink *et al.,* "Nanophotonics: Shrinking light-based technologies", *Science* **348**, 516-521(2015).